# SPMF: A Social Trust and Preference Segmentation-based Matrix Factorization Recommendation Algorithm

Wei Peng,  Baogui Xin *

College of Economics and Management, Shandong University of Science and Technology, Qingdao 266590, China

*Corresponding author

E-mail address: pengweisd@foxmail.com (W. Peng), xin@tju.edu.cn (B. Xin)

## ABSTRACT

The traditional social recommendation algorithm ignores the following fact: the preferences of users with trust relationships are not necessarily similar, and the consideration of user preference similarity should be limited to specific areas. A social trust and preference segmentation-based matrix factorization (SPMF) recommendation system is proposed to solve the above-mentioned problems. Experimental results based on the Ciao and Epinions datasets show that the accuracy of the SPMF algorithm is significantly higher than that of some state-of-the-art recommendation algorithms. The proposed SPMF algorithm is a more accurate and effective recommendation algorithm based on distinguishing the difference of trust relations and preference domain, which can support commercial activities such as product marketing.

## INTRODUCTION

The Internet has brought about industrial change and nurtured e-commerce. E-commerce has generated huge amount of network information, which results in information overload. Information overload directly increases the difficulty of selecting products and inspires people to seek effective solutions. Today, there are mainly four types of solutions we can employ: (1) information acquisition timelines; (2) categories; (3) search engines; and (4) personalized recommendations.

The first type of solution can save information retrieval time, but it is easy to miss lots of useful information. The second type of solution is to classify the project according to the similarity feature chosen by the user; this can overcome the defects of the first type of solution, but it has the disadvantage of low efficiency and poor precision. The third type of solution allows users to retrieve and filter irrelevant information by keywords, which can solve the problem of the second type of solution but cannot consider the user's individualized needs. Finally, the fourth type of solution can solve the problem of the third solution, and through historical data, user attributes, product attributes, and other information to mine user preferences, it can actively make personalized recommendations to users (Kuang et al., 2018).

A recommendation system can inspire the potential needs of users and make an e-commerce platform more intelligent and humanized. Such systems have helped Amazon, JD, Alibaba, and other companies to significantly increase sales. The recommendation algorithm is the core of the recommendation system (Chang, Lin, & Chen, 2016), but data sparsity and other problems have always been obstacles to its further development. Most scholars (Khazaei & Alimohammadi, 2018) employ machine learning algorithms, such as those for clustering and dimensionality reduction, to fill sparse data with a small amount of original data. However, it is difficult to guarantee the quality of the filled data.

In order to solve the above problems and improve the accuracy of recommendation, many scholars (AlBanna, Sakr, Moussa, & Moawad, 2016; Cui et al., 2018; Haruna et al., 2017) put forward recommendation algorithms based on the social network, that is, using direct or indirect trust relationships to make recommendations for target users. However, the trust relationship-based algorithm still has three problems: (1) The preference similarity of users with the trust relationship may be small or even zero. For example, a user has a trust relationship with his parents, but their preference differences may be significant, and therefore the recommendation may not be ideal. (2) The diversity of user preference determines that the measurement of preference similarity should be limited to certain areas. For example, the preference similarity of users is low in the music field, but it may be very high in the film field, and therefore a recommendation in the film field is very satisfactory. (3) Even if the preferences of different users are very similar, the recommendation will significantly influenced because of the difference in trust relationships.

Based on the above discussions, our main contributions are as follows. We proposed a social trust and preference segmentation-based matrix factorization recommendation algorithm by setting different recommendation trust weights for different relationships and the preference domain. This solves the above-mentioned problems and improves the recommendation accuracy.

*Figure 1. The context of the literature in this paper*

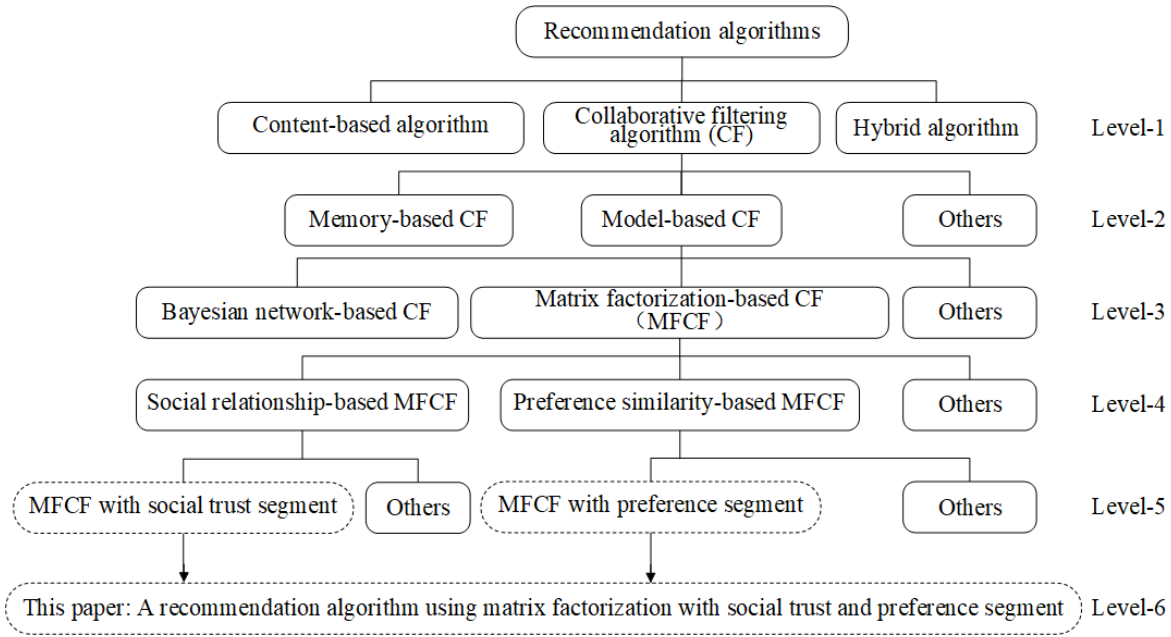

## BACKGROUND

The context of the literature in this paper is shown in Figure 1. As the level-1 classification shows, Cui et al. (2018) and Su & Khoshgoftaar (2009) divided the recommendation algorithms into three categories: content-based recommendation algorithms (Pazzani & Billsus, 2007), collaborative filtering algorithms (Herlocker, Konstan, & Riedl, 2000), and hybrid recommendation algorithms (Burke, 2002). Strictly speaking, the content-based recommendation algorithm is derived from the collaborative filtering recommendation algorithm, and both generate recommendations for target users based on similarity. Content-based recommendation algorithms need to filter massive amounts of information and update user profiles regularly, resulting in high time complexity and unsatisfactory recommendation results. Conversely, collaborative filtering recommendation algorithms only use neighbors with high similarity to target users to evaluate a product, predict the preferences of target users, and make recommendations. Hybrid recommendation algorithms are designed to meet a specific need and incorporate content-based recommendation algorithms and collaborative filtering recommendation algorithms.

The collaborative filtering algorithm is the most successful and widely used personalized recommendation algorithm in business, and it is also a hotspot of academic research. As the level-2 classification shows, Breese et al. (1998) divided this algorithm into a memory-based collaborative filtering algorithm and a model-based collaborative filtering algorithm. The memory-based collaborative filtering algorithm does not distinguish the rated item information attributes, and therefore it directly uses the correlation matrix to make predictions, which results in a heavy workload and low efficiency. In order to improve the recommendation efficiency and accuracy, the model-based collaborative filtering algorithm employs machine learning and data mining models such as the Bayesian network (Kang, 2018), SVM (Ren & Wang, 2018), or matrix decomposition (Zhao & Sun, 2018) for recommendation; this is shown in the level-3 classification. However, the ability to solve the sparsity problem of user-item rating data is an important indicator for evaluating the pros and cons of a recommendation algorithm. The matrix factorization-based collaborative filtering algorithm has become one of the most popular algorithms in the last decade due to its outstanding performance in the 2009 Netflix Prize competition.

As the level-4 classification shows, the matrix factorization-based collaborative filtering algorithm includes the trust relationship-based matrix factorization algorithm (Ma, Yang, Lyu, & King, 2008; Mi, Peng, Xiao, & Lu, 2017) and the preference similarity-based matrix factorization algorithm (Han, Huang, Zhang, & Bhatti, 2018). Wang et al. (2006) solved the problem of low-accuracy caused by sparse data by integrating the user preference similarity and matrix factorization algorithm and combining the user rating data. Lai et al. (2019) constructed a social recommendation model that integrated trust relationships and product popularity and speculated on their potential interactions based on user interaction behavior. The trust relationship could lead to more accurate recommendations. Guo et al. (2016) proposed TrustSVD, a trust-based matrix factorization technique for recommendations. Lee & Ma (2016) constructed a recommendation algorithm that combined the KNN and matrix factorization by using a trust relationship and propagation effect, and combining user rating and the trust relationship, they achieved a higher recommendation accuracy.

In this paper, we will combine the social trust segmentation and the preference domain segmentation and construct the level-6 classification, i.e., a social trust and preference segmentation-based matrix factorization (SPMF) recommendation algorithm.

## RESEARCH MODEL

The main ideas of the proposed SPMF recommendation algorithm are as follows:
(1) To show the impact of preference domain segmentation on target user recommendation.
(2) To reveal the different influences on target users' recommendation between users with and without trust relationships.

### Preference Domain Segmentation

User preferences are diverse, but most of the existing recommendation algorithms do not take full account of it. To demonstrate the impact of preference domain segmentation on measuring user preference similarity, we provide the following example.

A website provides ten products that can be subdivided into three categories: music, movies, and books (Table 1). The numbers in the table represent rating records. The rating scale is 1–5, and 0 indicates no rating. $u_a$ and $u_b$ represent users, and $I_k^P$ denotes the $k^{th}$ item in the domain $P$.

*Table 1. Rating records of a website*

| Items<br>Users | Music | | | | Movies | | | Books | | |
|---|---|---|---|---|---|---|---|---|---|---|
| | $I_1^1$ | $I_2^1$ | $I_3^1$ | $I_4^1$ | $I_1^2$ | $I_2^2$ | $I_3^2$ | $I_1^3$ | $I_2^3$ | $I_3^3$ |
| $u_a$ | 4 | 2 | 5 | 0 | 0 | 0 | 0 | 5 | 4 | 5 |
| $u_b$ | 4 | 1 | 5 | 5 | 3 | 5 | 1 | 1 | 0 | 2 |

Table 1 shows that the preferences of $u_a$ and $u_b$ in the music field are similar; the preferences are temporarily uncertain in the movie field; and the preferences are quite different in the book field.

We measure preference similarities $sim(u_a, u_b)^{COS}$, $sim(u_a, u_b)^{PCC}$, and $sim(u_a, u_b)^{Jaccard}$ of users $u_a$ and $u_b$ by using classical methods COS (Salah, Rogovschi, & Nadif, 2016), PCC (Salah, Rogovschi, & Nadif, 2016), and Jaccard (Liu, Hu, Mian, Tian, & Zhu, 2014), respectively. Then, we compare them with the similarity after preference domain segmentation. This comparison is shown in Table 2.

*Table 2. Comparison of preference similarity*

| Measures / Values / Domains | Without preference domain segmentation | | | With preference domain segmentation | | |
|---|---|---|---|---|---|---|
| | $sim(u_a,u_b)^{COS}$ | $sim(u_a,u_b)^{Jaccard}$ | $sim(u_a,u_b)^{PCC}$ | $ssim(u_a,u_b)^{COS}$ | $ssim(u_a,u_b)^{Jaccard}$ | $ssim(u_a,u_b)^{PCC}$ |
| Music | 0.53 | 0.50 | 0.29 | 0.78 | 0.75 | 0.97 |
| Movies | | | | -- | -- | -- |
| Books | | | | 0.82 | 0.67 | 0.00 |

Table 2 shows that the preference similarity with preference domain segmentation can describe preference similarity among users more precisely than that without preference domain segmentation. Furthermore, PCC is the most accurate method. Moreover, from $ssim(u_a, u_b)^{PCC}$ we can predict that $u_a$ may be interested in $I_4^1$. In this paper, we will employ $ssim(u_a, u_b)^{PCC}$ to calculate the preference similarity. The formula is as follows:

$$ssim(u_a, u_b)^{PCC} = \frac{\sum_{i \in I_{u_a}^P \cap I_{u_b}^P} \left(r_{u_a,i} - \overline{r}_{u_a}\right)\left(r_{u_b,i} - \overline{r}_{u_b}\right)}{\sqrt{\sum_{i \in I_{u_a}^P \cap I_{u_b}^P} \left(r_{u_a,i} - \overline{r}_{u_a}\right)^2} \sqrt{\sum_{i \in I_{u_a}^P \cap I_{u_b}^P} \left(r_{u_b,i} - \overline{r}_{u_b}\right)^2}}, \quad (1)$$

where $I_{u_a}$ and $I_{u_b}$ represent the rated item sets of $u_a$ and $u_b$, respectively; $r_{u_a,i}$ and $r_{u_b,i}$ indicate the ratings of $u_a$ and $u_b$ for a specific item $i$, respectively; and $\overline{r}_{u_a}$ and $\overline{r}_{u_b}$ denote the average rating of $u_a$ and $u_b$, respectively.

Considering the preference domain segmentation of users, we can measure a user's experience in a particular domain based on the number of items bought by the user. The formula is as follows:

$$\xi_u^P = \frac{|I_u|}{|I_{u_1} \cup I_{u_2} \cup \cdots \cup I_{u_m}|}, \tag{2}$$

where $I_u$ denotes the item set rated by the target user $u$; $|\cdot|$ indicates the element amount of the set; and $u_1, u_2, \ldots, u_m$ represent all users in the particular domain $P$.

*Figure 2. Schematic diagram of the process of the SPMF recommendation algorithm (A: PMF, B: SocialMF, C: STE, D: SPMF).*

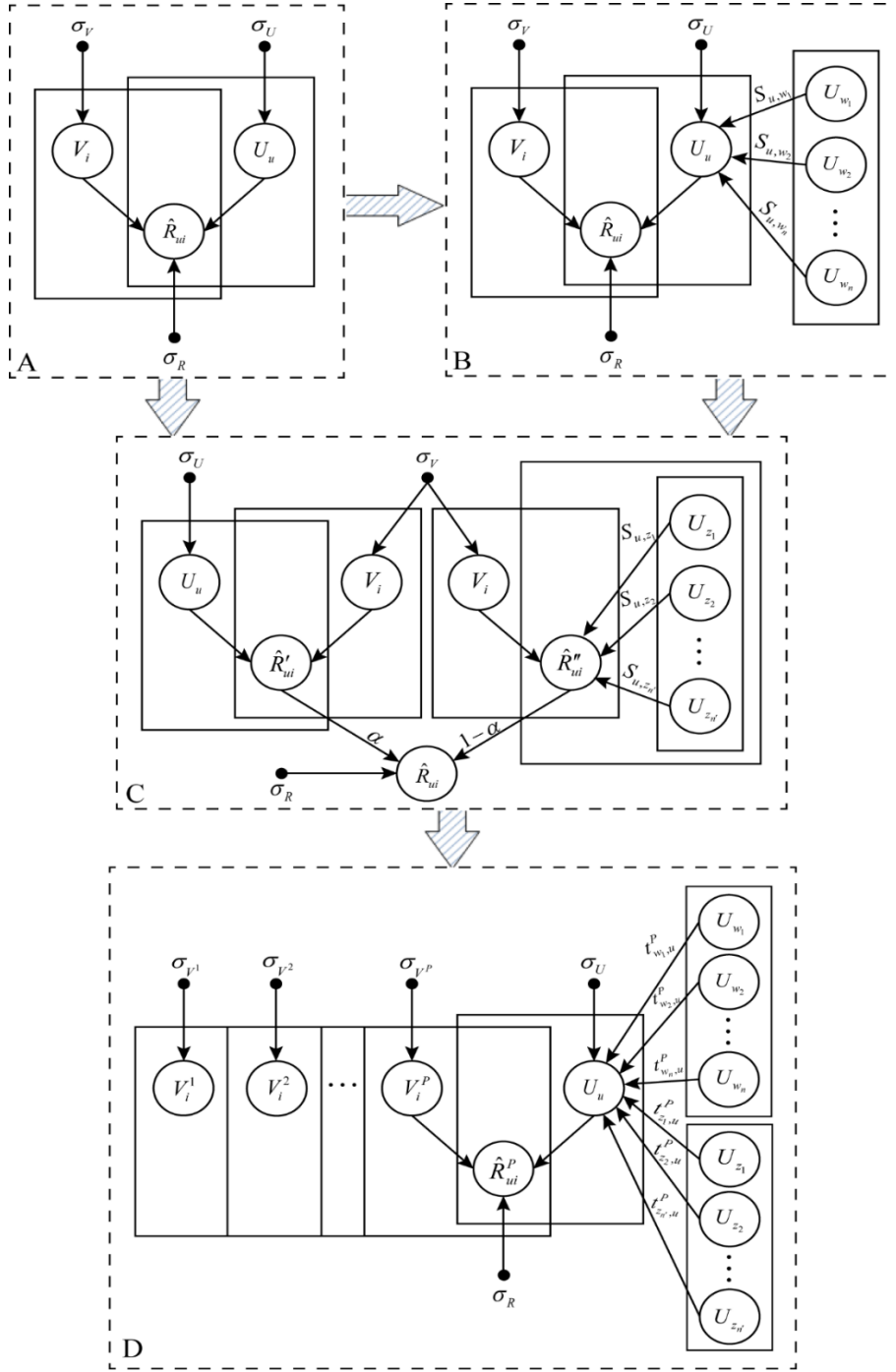

## Trust Relationship Segmentation

In Figure 2, the user eigenmatrix $U_u$ obeys the Gaussian distribution with the mean $\mu = 0$ and the variance $\sigma^2 = \sigma_U^2$; the item eigenmatrix $V_i$ obeys the Gaussian distribution with the mean $\mu = 0$ and the variance $\sigma^2 = \sigma_V^2$; and the predicted rating $\hat{R}_{ui}$ obeys the Gaussian distribution with the mean $\mu = r_{ui}$ and the variance $\sigma^2 = \sigma_R^2$. $S_{u_a,u_b}$ denotes the social link value, where $S_{u_a,u_b} = 0$ indicates no social relationship between $u_a$ and $u_b$. $L_u$ represents the set of users who have a social link with the target user $u$, and $w_1, w_2, \cdots w_n$ denote elements of $L_u$. $N_u$ represents the set of users who do not have a social link with the target user $u$, and $z_1, z_2, \cdots z_{n'}$ denote elements of $N_u$. $t_{w,u}^P$ and $t_{z,u}^P$ denote different recommendation influences in specific domain $P$, which can be calculated by Eq. (3).

User relationships are complex, but in classical recommendation algorithms, we assume that user relationships are independent of each other in order to simplify the problem. Typically, we only consider the user's ratings on products, such as with the probability matrix factorization (PMF) recommendation algorithm (Jamali & Ester, 2010), as shown in Figure 2(A). However, it is difficult for us to be completely independent when making sensible decisions, and we are often influenced by our friends and family. In view of this, some researchers introduced the trust relationship into the recommendation algorithm and only considered the user evaluation with a trust relationship to make recommendations for target users, i.e., the SocialMF recommendation algorithm (Jamali & Ester, 2010). This algorithm is shown in Figure 2(B). Social recommendation algorithms could improve recommendation accuracy but ignored the user's evaluation of the product. In order to overcome the shortcomings of the PMF recommendation algorithm and the trust-based recommendation algorithm and utilize their advantages, it is necessary to combine them. Doing so produces the STE recommendation algorithm (Ma, King, & Lyu, 2009), as shown in Figure 2(C). In order to adequately reflect the effect of the trust relationship and the preference similarity to target users, we construct a matrix factorization recommendation algorithm that combines trust relationship segmentation and preference domain segmentation, named SPMF, which is shown in Figure 2(D).

In order to distinguish the different influences of $L_u$ and $N_u$ on $u$, we define the specific domain recommendation influence $t_{u',u}^P$ as follows:

$$t_{u',u}^P = \begin{cases} \alpha \times ssim(u',u)^{PCC} \times \xi_{u'}^P, & u' \in L_u, \\ (1-\alpha) \times ssim(u',u)^{PCC} \times \xi_{u'}^P, & u' \in N_u, \end{cases} \qquad (3)$$

where $\alpha$ is an adjustment factor that is used to weigh the recommendation impact of $L_u$ and $N_u$ on the target user $u$; and $u'$ represents another user who belongs to either $L_u$ or $N_u$.

In the specific domain $P$, we can define the predicted target user eigenvector $\hat{U}_u$ according to the eigenvectors of $L_u$ and $N_u$:

$$\hat{U}_u = \sum_{w \in L_u} t_{w,u}^P U_w + \sum_{z \in N_u} t_{z,u}^P U_z . \qquad (4)$$

If we incorporate the specific domain recommendation influence into the PMF algorithm, the posterior probability distribution of the user eigenmatrix $U$ and the item eigenmatrix $V$ is as follows:

$$p\left(U,V \mid R,T,\sigma_R^2,\sigma_T^2,\sigma_U^2,\sigma_V^2\right) \propto p\left(R \mid U,V,\sigma_R^2\right) p\left(U \mid T,\sigma_U^2,\sigma_T^2\right) p\left(V \mid \sigma_V^2\right)$$

$$= \prod_{u=1}^{m} \prod_{i=1}^{n} \left[ N\left(r_{ui} \mid U_u^T V_i, \sigma_R^2\right) \right]^{\delta_{ui}}$$

$$\times \prod_{u=1}^{n} N\left(U_u \mid \sum_{w \in L_u} t_{w,u}^P U_w + \sum_{z \in N_u} t_{z,u}^P U_z, \sigma_T^2 I\right) \quad , \quad (5)$$

$$\times \prod_{u=1}^{m} N(U_u \mid 0, \sigma_U^2 I) \times \prod_{i=1}^{n} N(V_i \mid 0, \sigma_V^2 I)$$

where $N(x \mid \mu, \sigma^2)$ denotes a Gaussian distribution with a mean $\mu$ and a variance $\sigma^2$; $\delta_{ui}$ is a coefficient in which $\delta_{ui} = 1$ means that $u$ has rated $i$, while $\delta_{ui} = 0$ represents no rating record; and $I$ is a unit matrix of $K$ dimensions.

The objective function of this algorithm is obtained by taking a negative logarithm of Eq. (5) as follows:

$$L(R,T,U,V) = \frac{1}{2} \sum_{u=1}^{m} \sum_{i=1}^{n} I_{ui} \left(r_{ui} - U_u^T V_i\right)^2 + \frac{\lambda_u}{2} \sum_{u=1}^{m} U_u^T U_u + \frac{\lambda_v}{2} \sum_{i=1}^{n} V_i^T V_i$$

$$+ \frac{\lambda_t}{2} \sum_{u=1}^{m} \left( \left( U_u - \left( \sum_{w \in L_u} t_{w,u}^P U_w + \sum_{z \in N_u} t_{z,u}^P U_z \right) \right)^T \left( U_u - \left( \sum_{w \in L_u} t_{w,u}^P U_w + \sum_{z \in N_u} t_{z,u}^P U_z \right) \right) \right) + C \quad , \quad (6)$$

where $\lambda_u = \frac{\sigma_R^2}{\sigma_U^2}$, $\lambda_v = \frac{\sigma_R^2}{\sigma_V^2}$, $\lambda_t = \frac{\sigma_R^2}{\sigma_T^2}$, and $C$ is constant.

In order to obtain the optimal value of the objective function (6), we employ the gradient descent algorithm to get the partial derivatives of the eigenvectors $U_u$ and $V_i$:

$$\frac{\partial L}{\partial U_u} = \sum_{i=1}^{n} I_{ui} V_i \left(U_u^T V_i - r_{ui}\right) + \lambda_u U_u + \lambda_t \left[ U_u - \left( \sum_{w \in L_u} t_{w,u}^P U_w + \sum_{z \in N_u} t_{z,u}^P U_z \right) \right]$$

$$- \lambda_t \sum_{u \in L_w} t_{u,w}^P \left[ U_w - \left( \sum_{a \in L_w} t_{a,w}^P U_a + \sum_{b \in N_w} t_{b,w}^P U_b \right) \right] \quad , \quad (7)$$

$$- \lambda_t \sum_{u \in N_z} t_{z,u}^P \left[ U_z - \left( \sum_{c \in L_z} t_{c,z}^P U_c + \sum_{d \in N_z} t_{d,z}^P U_d \right) \right]$$

$$\frac{\partial L}{\partial V_i} = \sum_{u=1}^{m} I_{ui} U_u \left(U_u^T V_i - r_{ui}\right) + \lambda_v V_i . \quad (8)$$

In this way, we can get the following eigenvectors by iteratively updating the gradient descent algorithm:

$$U_u^{(\tau+1)} = U_u^{(\tau)} + \gamma \begin{bmatrix} \sum_{i=1}^{n} I_{ui} V_i \left( r_{ui} - U_u^T V \right) - \lambda_u U_u - \lambda_t \left[ U_u - \left( \sum_{w \in L_u} t_{w,u}^P U_w + \sum_{z \in N_u} t_{z,u}^P U_z \right) \right] \\ -\lambda_t \sum_{u \in L_w} t_{u,w}^P \left[ U_w - \left( \sum_{a \in L_w} t_{a,w}^P U_a + \sum_{b \in N_w} t_{b,w}^P U_b \right) \right] \\ -\lambda_t \sum_{u \in N_z} t_{z,u}^P \left[ U_z - \left( \sum_{c \in L_z} t_{c,z}^P U_c + \sum_{d \in N_z} t_{d,z}^P U_d \right) \right] \end{bmatrix}, \qquad (9)$$

$$V_i^{(\tau+1)} = V_i^{(\tau)} + \gamma \left[ \sum_{u=1}^{m} I_{ui} U_u \left( r_{ui} - U_u^T V_i \right) - \lambda_v V_i \right], \qquad (10)$$

where $\tau$ is the sequence number of the iterative updating, and $\gamma$ is the learning rate.

## DATA ANALYSES AND RESULTS

### Experimental Datasets

In this section, two public datasets, Ciao and Epinions, are employed. The performance of the proposed SPMF is evaluated by the mean absolute error (MAE) and the root mean square error (RMSE) (Jamali & Ester, 2010; Ma et al., 2009). The statistics of datasets Ciao and Epinions are shown in Table 3. Obviously, both datasets are highly sparse.

*Table 3. Statistics of the experimental datasets*

|  | **Ciao** | **Epinions** |
| --- | --- | --- |
| Number of users ($m$) | 7375 | 40,163 |
| Number of items ($n$) | 99,746 | 139,738 |
| Number of rating records | 280,391 | 664,824 |
| Number of trust relationships | 111,781 | 487,183 |
| Sparsity of rating records | 99.9619% | 99.9882% |
| Sparsity of trust relationships | 99.7945% | 99.9698% |

### Algorithm Implementations

#### 1. Generate a Recommendation Influence Matrix of Specific Domains

For the SPMF recommendation algorithm, the recommendation influence matrix element of specific domains can be calculated by Eq. (3), and the algorithm generated by the recommendation influence matrix $T^{m \times m}$ of specific domains is shown by Algorithm 1.

**Algorithm 1.** Generate a recommendation matrix $T^{m \times m}$ of specific domains

| **Input:** the rating matrix $R^{m \times n}$ of the user-item, the original trust relationship matrix $S^{m \times m}$ among users |
| --- |
| **Output:** the recommendation influence matrix $T^{m \times m}$ of specific domains |

| | |
|---|---|
| 1 | **Procedure** Rec_Influence(R, S, alpha): |
| 2 | Count = []; |
| 3 | **for** *i* in *sum_users* **do** |
| 4 | *count_1* = 0; *R_sum* =0; *R_avg* =[]; |
| 5 | **for** *j* in *sum_items* **do** |
| 6 | **if** *R[i][j]*>0 **then** |
| 7 | *count_1* = *count_1* + 1; Count.append(*count_1*) |
| 8 | *R_sum* = *R_sum*+ *R[i][j]*; |
| 9 | **end if** |
| 10 | *R_avg.append* (*count_1*/ *R[i]_sum*); |
| 11 | **end for** |
| 12 | **end for** |
| 13 | *e* = [] |
| 14 | **for** *u* in *sum_users* **do** |
| 15 | *count_2* = 0; |
| 16 | **for** *v* in *sum_users* **do** |
| 17 | *S_c* = 0; *S_u* = 0; *S_v* = 0; |
| 18 | **if** (*S[u][v]*==1): |
| 19 | *alpha[u][v]* ==0.4; |
| 20 | **else**: |
| 21 | *alpha[u][v]* ==0.6; |
| 22 | **end if** |
| 23 | **for** *j* in *sum_items* **do** |
| 24 | **if** ((*R[u][j]*>0) or (*R[v][j]*>0)) **then** |
| 25 | count_2 = count_2 + 1; |
| 26 | *S_c* = *S_c* + (*R[u][j]* - *R _avg[u]*)* (*R[v][j]* - *R _avg[v]*); |
| 27 | *S_u* = *S_u* + (*R[u][j]* - *R_avg[u]*)* (*R[u][j]* - *R _avg[u]*); |
| 28 | *S_v* = *S_v* + (*R[v][j]* - *R_avg[v]*)* (*R[v][j]* - *R _avg[v]*); |
| 29 | **end if** |
| 30 | **end for** |
| 31 | *s[u][v]* = *S_c/((sqrt(S_u))* (sqrt(S_v)))*; |
| 32 | **end for** |
| 33 | *e[u]* = (*Count[u]*/float(*count_2*)); |
| 34 | *t[u][v]* = *alpha[u][v]*s[u][v]* e[u]*; |
| 35 | **end Procedure**; |

In Algorithm 1, $\alpha$ is set to 0.4. We explain this in the section titled "Parameter analysis on $\alpha$."

2. **The User-Item Rating Matrix for Prediction**

For the SPMF recommendation algorithm, the algorithm of the user-item rating matrix for prediction is shown by Algorithm 2.

**Algorithm 2.** The user-item rating matrix for prediction

**Input:** the rating matrix $R^{m \times n}$, the recommendation influence matrix $T^{m \times m}$ of specific domains

**Output:** the predicted rating matrix $\hat{R}^{m \times n}$

| 1  | **for** *epo* in *epochs* **do** |
|----|---|
| 2  |   **for** *i* in *sum_users* **do** |
| 3  |     **for** *j* in *sum_items* **do** |
| 4  |       **if** *R[i][j]>0* **then** |
| 5  |         *eij=R[i][j]-np.dot(P[i,:],Q[:,j])*; |
| 6  |       **end if** |
| 7  |       **for** *k* in *K* **do** |
| 8  |         **for** *m* in *sum_users* **do** |
| 9  |         **for** *n* in *sum_users* **do** |
| 10 |           *P[i][k]=P[i][k]+lr\*(eij\*Q[k][j]-lambda_u_v\*P[i][k]-lambda_t\*(P[i][k]-t[i][m]\*P[m][k])+lambda_t\*T[i][m]\*(P[m][k]-t[m][n]\*P[n][k]))*; |
| 11 |           *Q[k][j]=Q[k][j]+lr\*(eij\*P[i][k]-lambda_u_v\*Q[k][j])*; |
| 12 |         **end for** |
| 13 |         **end for** |
| 14 |       **end for** |
| 15 |     **end for** |
| 16 |   **end for** |
| 17 |   *R_MF=np.dot(P,Q)*; |
| 18 | **end for** |

In Algorithm 2, $K$ represents matrix factorization dimensions.

## Experimental Results Comparison

For comparison, we use the same strategy as most of the literature: the matrix factorization dimensions are set to $K = 5$ and $10$. The MAE and RMSE of the proposed SPMF algorithm and other recommendation algorithms on Epinions and Ciao datasets are shown in Table 4.

*Table 4. RMSE comparison with baselines*

| Datasets | Dimension | Indicator | PMF | STE | SocialMF | TrustSVD | SPMF | Accuracy improvement |
|---|---|---|---|---|---|---|---|---|
| Epinions | K=5 | MAE | 0.979 | 0.950 | 0.825 | 0.804 | **0.794** | 1.24% |
| | | RMSE | 1.290 | 1.196 | 1.070 | 1.043 | **0.989** | 5.18% |
| | K=10 | MAE | 0.909 | 0.958 | 0.826 | 0.805 | **0.762** | 5.34% |
| | | RMSE | 1.197 | 1.278 | 1.082 | 1.044 | **0.974** | 6.70% |
| Ciao | K=5 | MAE | 0.920 | 0.767 | 0.749 | 0.723 | **0.571** | 21.02% |
| | | RMSE | 1.260 | 1.020 | 0.981 | 0.955 | **0.759** | 20.52% |
| | K=10 | MAE | 0.822 | 0.763 | 0.749 | 0.723 | **0.565** | 21.85% |
| | | RMSE | 1.078 | 1.013 | 0.976 | 0.956 | **0.756** | 20.92% |

Table 4 shows that the MAE and RMSE of the SPMF algorithm are less than those of algorithms PMF, RSTE, SocialMF, and TrustSVD in the cases of $K=5$ and $10$, i.e., the accuracy of the SPMF algorithm is the highest among all compared algorithms.

## Parameter Analysis

In exploring the influence of various parameters on the performance of the algorithm, this paper adopts the idea of control variables. The more the epoch increases, the longer the training time becomes. Therefore, the epochs in this section are all set at 10.

1. **Parameter Analysis on $\lambda_u$ and $\lambda_v$**

In this section, we explore the effect of $\lambda_u$ and $\lambda_v$ values on algorithm performance under different decomposition dimensions. The results are shown in Figures 3 and 4.

*Figure 3. RMSE vs. $\lambda_u$ and $\lambda_v$ on Ciao*

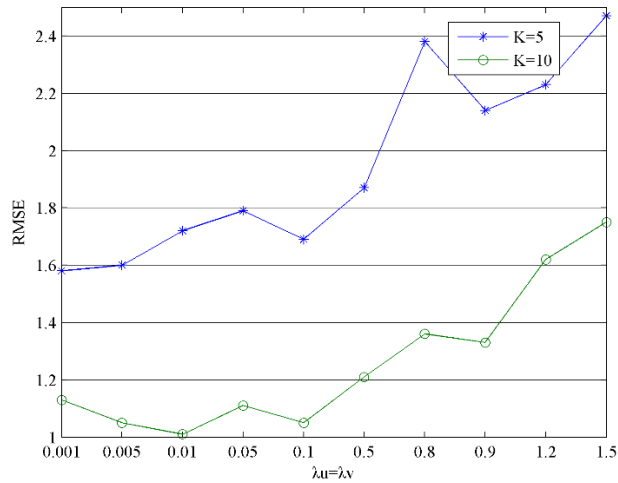

*Figure 4. RMSE vs. $\lambda_u$ and $\lambda_v$ on Epinions*

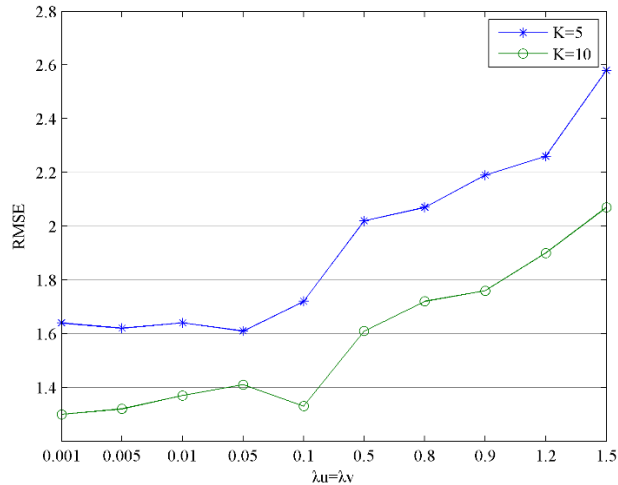

Figures 3 and 4 show that the optimal parameter values should be $\lambda_u = \lambda_v = 0.005$ on both Ciao and Epinions since the average RMSE corresponding to different decomposition dimensions is the smallest only when $\lambda_u = \lambda_v = 0.005$.

2. **Parameter Analysis on $\lambda_t$**

In this section, we explore the effect of $\lambda_t$ values on algorithm performance under different decomposition dimensions. The results on datasets Ciao and Epinions are shown in Figures 5 and 6. The different $\lambda_t$ values are set as in the previous literature.

*Figure 5. RMSE vs. $\lambda_t$ on Ciao*

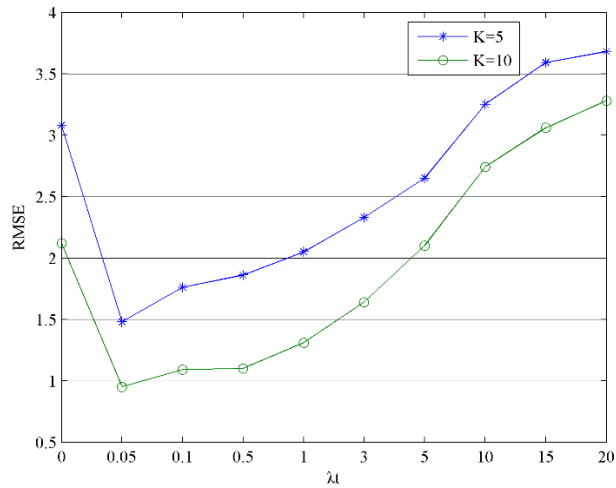

*Figure 6. RMSE vs. $\lambda_t$ on Epinions*

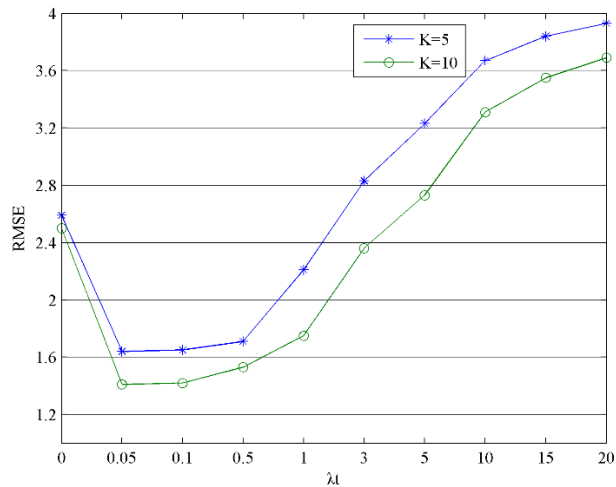

From Figures 5 and 6, we can conclude that in different decomposition dimensions of different datasets, the RMSE value of the algorithm is the lowest when $\lambda_t=0.05$ on both datasets.

3. **Parameter Analysis on** $\alpha$

In this section, we explore the optimal $\alpha$ of the algorithm. The result is shown in Figure 7.

*Figure 7. RMSE vs. $\alpha$ on Ciao and Epinions*

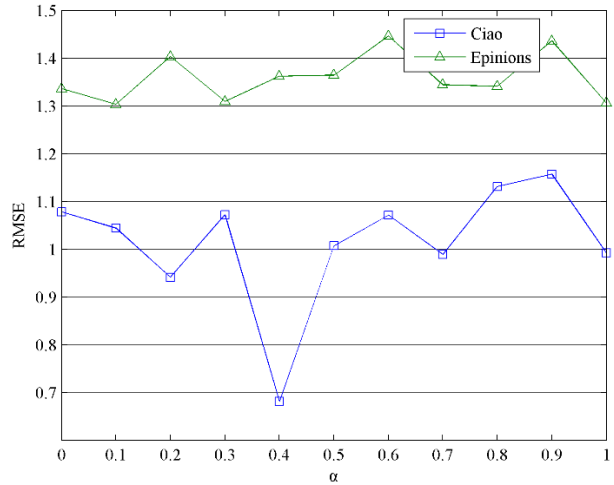

Figure 7 shows that the optimal parameter value should be $\alpha=0.4$ on datasets Ciao and Epinions because the average RMSE is the smallest when $\alpha=0.4$.

4. **Parameter Analysis on** $K$

In this section, we explore the optimal decomposition dimension $K$. The result is shown in Figure 8.

*Figure 8. RMSE vs. $K$ on Ciao and Epinions*

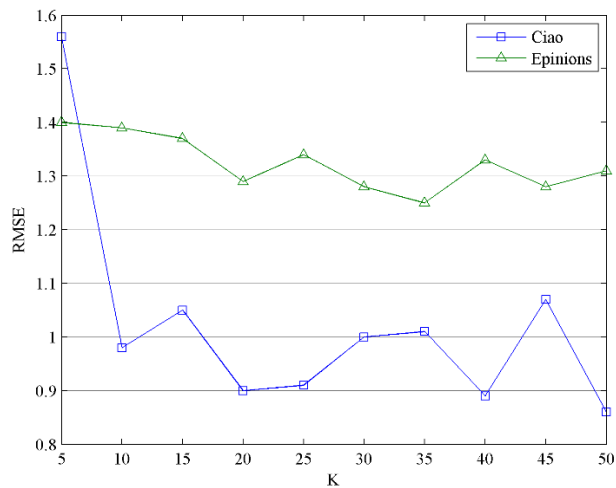

Figure 8 shows that the optimal factorization dimension of both datasets is $K=20$. On Ciao, the RMSE of $K=20$ is not the smallest in this experiment, but too large of a decomposition dimension may cause overfitting, and the difference of RMSE between $K=20$ and $K=40$, $K=50$ is within a reasonable range, and. Similarly, we can conclude that $K=20$ is the optimal decomposition dimension of Epinions.

## CONCLUSIONS

The traditional social recommendation algorithm ignores the following fact: the preferences of users with trust relationships are not necessarily similar, and the consideration of user preference similarity should be limited to specific areas. To solve these problems, this paper proposed a SPMF recommendation algorithm by setting different recommendation trust weights for different relationships and preference domains. The experimental results on Ciao and Epinions datasets show that the accuracy of the SPMF algorithm is much higher than that of some state-of-the-art recommendation algorithms.

## ACKNOWLEDGMENT

The authors would like to express sincere gratitude to the referees for their valuable suggestions and comments. The work is supported by the National Social Science Foundation of China [No. 16FJY008]; the Natural Science Foundation of Shandong Province [No. ZR2016FM26].

## REFERENCES

AlBanna, B., Sakr, M., Moussa, S., & Moawad, I. (2016). Interest Aware Location-Based Recommender System Using Geo-Tagged Social Media. *ISPRS International Journal of Geo-Information, 5*(12), 245.

Breese, J. S., Heckerman, D., & Kadie, C. (1998). *Empirical analysis of predictive algorithms for collaborative filtering.* Paper presented at the Proceedings of the Fourteenth conference on Uncertainty in artificial intelligence.

Burke, R. (2002). Hybrid recommender systems: Survey and experiments. *User modeling and user-adapted interaction, 12*(4), 331-370.

Chang, P.-C., Lin, C.-H., & Chen, M.-H. (2016). A hybrid course recommendation system by integrating collaborative filtering and artificial immune systems. *Algorithms, 9*(3), 47.

Cui, L., Huang, W., Yan, Q., Yu, F. R., Wen, Z., & Lu, N. (2018). A novel context-aware recommendation algorithm with two-level SVD in social networks. *Future Generation Computer Systems, 86*, 1459-1470.

Guo, G., Zhang, J., & Yorke-Smith, N. (2016). A novel recommendation model regularized with user trust and item ratings. *ieee transactions on knowledge and data engineering, 28*(7), 1607-1620.

Han, H., Huang, M., Zhang, Y., & Bhatti, U. (2018). An Extended-Tag-Induced Matrix Factorization Technique for Recommender Systems. *Information, 9*(6), 143.

Haruna, K., Akmar Ismail, M., Suhendroyono, S., Damiasih, D., Pierewan, A., Chiroma, H., & Herawan, T. (2017). Context-aware recommender system: A review of recent developmental process and future research direction. *Applied Sciences, 7*(12), 1211.

Herlocker, J. L., Konstan, J. A., & Riedl, J. (2000). *Explaining collaborative filtering recommendations.* Paper presented at the Proceedings of the 2000 ACM conference on Computer supported cooperative work.

Jamali, M., & Ester, M. (2010). *A matrix factorization technique with trust propagation for recommendation in social networks.* Paper presented at the Proceedings of the fourth ACM conference on Recommender systems.

Kang, S. (2018). Outgoing call recommendation using neural network. *Soft Computing, 22*(5), 1569-1576.

Khazaei, E., & Alimohammadi, A. (2018). An Automatic User Grouping Model for a Group Recommender System in Location-Based Social Networks. *ISPRS International Journal of Geo-Information, 7*(2), 67.

Kuang, L., Yu, L., Huang, L., Wang, Y., Ma, P., Li, C., & Zhu, Y. (2018). A Personalized QoS Prediction Approach for CPS Service Recommendation Based on Reputation and Location-Aware Collaborative Filtering. *Sensors, 18*(5), 1556.

Lai, C.-H., Lee, S.-J., & Huang, H.-L. (2019). A social recommendation method based on the integration of social relationship and product popularity. *International Journal of Human-Computer Studies, 121*, 42-57.

Lee, W.-P., & Ma, C.-Y. (2016). Enhancing collaborative recommendation performance by combining user preference and trust-distrust propagation in social networks. *Knowledge-Based Systems, 106*, 125-134.

Liu, H., Hu, Z., Mian, A., Tian, H., & Zhu, X. (2014). A new user similarity model to improve the accuracy of collaborative filtering. *Knowledge-Based Systems, 56*, 156-166.

Ma, H., King, I., & Lyu, M. R. (2009). *Learning to recommend with social trust ensemble.* Paper presented at the Proceedings of the 32nd international ACM SIGIR conference on Research and development in information retrieval.

Ma, H., Yang, H., Lyu, M. R., & King, I. (2008). *Sorec: social recommendation using probabilistic matrix factorization.* Paper presented at the Proceedings of the 17th ACM conference on Information and knowledge management.

Mi, C., Peng, P., Xiao, L., & Lu, Y. (2017). *Recommendation Algorithm Based on User Trust and Interest with Probability Matrix Factorization.* Paper presented at the 2017 Fifth International Conference on Advanced Cloud and Big Data (CBD).

Pazzani, M. J., & Billsus, D. (2007). Content-based recommendation systems *The adaptive web* (pp. 325-341): Springer.

Ren, L., & Wang, W. (2018). An SVM-based collaborative filtering approach for Top-N web services recommendation. *Future Generation Computer Systems, 78*, 531-543.

Salah, A., Rogovschi, N., & Nadif, M. (2016). A dynamic collaborative filtering system via a weighted clustering approach. *Neurocomputing, 175*, 206-215.

Wang, J., De Vries, A. P., & Reinders, M. J. (2006). *Unifying user-based and item-based collaborative filtering approaches by similarity fusion.* Paper presented at the Proceedings of the 29th annual international ACM SIGIR conference on Research and development in information retrieval.

Zhao, J., & Sun, G. (2018). Detect User's Rating Characteristics by Separate Scores for Matrix Factorization Technique. *Symmetry, 10*(11), 616.